\documentclass[letterpaper, 10 pt, conference]{ieeeconf}  

\IEEEoverridecommandlockouts                              
\usepackage{amsfonts}
\usepackage{mathtools}

\usepackage{bigints}
\usepackage{framed}
\usepackage{dirtytalk}
\usepackage{color}
\newcommand{\quotes}[1]{``#1''}
\usepackage{gensymb}
\usepackage{accents}

\usepackage{booktabs}
\usepackage{hyperref}
\usepackage{subfigure}
\usepackage{graphicx}

\newtheorem{theorem}{Theorem}
\newtheorem{lemma}{Lemma}

\title{\LARGE \bf
A Novel Framework for Decentralized Dynamic Resource Allocation Using Voronoi Tessellations
}

\author{Bhagyashri Telsang and Seddik Djouadi
\thanks{This paper was supported in part by the National Science Foundation under grant NSF-CMMI-2024111.}
\thanks{B. Telsang and S. Djouadi are with the Department of Electrical Engineering and Computer Science, University of Tennessee Knoxville, USA.
      {\tt\small \{btelsang,mdjouadi\}@utk.edu}}%
}

\begin{document}

\maketitle
\thispagestyle{empty}
\pagestyle{empty}

\begin{abstract}

In this work, we approach the problem of resource allocation in a team of agents through the framework of Centroidal Voronoi Tessellations. CVTs provide a natural way to embed a desired global trend in the team through probability distributions, and in one-dimensional spaces, CVTs offer an inherent line structure allowing for a simple communication graph and scalability. We first consider the amount of resource to be allocated to be a constant and provide an analytical solution to such static resource allocation problem by embedding the allocation constraint within the distribution through a system of nonlinear equations. Using the solution of such a constrained CVT minimization problem as an initialization step, we propose a decentralized dynamic resource allocation solution that employs a one-step update when the desired distribution is Gaussian. We introduce a \quotes{civility model} for negotiations between the agents to allow for flexibility in local preferences and maintaining robustness against local disturbances. We demonstrate the effectiveness of the proposed method by considering the application of demand-response in smart grids through the problem of power allocation in a group of building thermal loads.

\end{abstract}

\section{INTRODUCTION}
\label{sec::introduction}

Often times we see a conflicting, paradoxical problem around us in the world. There is too much and yet there is not enough; obesity and hunger coexisting, overpopulation and population scarcity coexisting, floods and drought coexisting few hundred miles apart, vacant house and homeless people outside of them. In each of the scenario, there is a resource that is abundant in one sector, be it location or a group of people, but scarce in another. It makes one wonder if we can alleviate the problem by allocating the resources in a \quotes{right} manner.

Taking roots in the field of Economics through \cite{GeneralCompetitiveAnalysis} in $1970's$, the resource allocation problem has broadened to the field of engineering in more recent decades. Mathematically, the resource allocation problem can be framed as, \cite{Hariharan}:

\begin{align}
	\min_{z_i \in \mathbb{R}^n } \frac{1}{N} \sum_{i \in I_N} f_i(z_i) \nonumber \\
	\text{such that,} \hspace{0.5cm} \sum_{i \in I_N} z_i = r
	\label{eq:resourceallocationproblemIntro}
\end{align}

 In the resource allocation problem, an $r$ amount of resource is to be allocated among $N$ agents while minimizing the sum of their individual cost functions $\{f_i\}_{i \in I_N}$. Typically, in engineering problems, the agents are local controllers tasked to maintain local interests while equipped with capabilities to communicate with other agents.
 
Simultaneously seeming trivial and complex, the nature of \eqref{eq:resourceallocationproblemIntro} can be broken down into the following aspects: the information structure in the group of agents, the separability of the objective function, and the global constraint. Due to the separability of the objective function, each agent can minimize the (global) cost function without any dependance on other agents. However, because of the global constraint imposed on the team, the team information structure becomes a significant aspect. 
 
Like most of the work on the resource allocation problem, the authors in \cite{Hariharan} assume the individual cost functions to be convex. In the case where the cost functions are differentiable, they propose a gradient descent consensus algorithm. And when the cost functions are not necessarily differentiable, they present a sub-gradient based algorithm. While they let the team information structure be dynamic, they impose reasonable mild conditions on the team information structure like connectedness, and start at an initial feasible condition. 

While \cite{Hariharan} proposes the gradient descent algorithm where the agents trade resources in proportion to the gradient difference for their individual cost functions, \cite{ResourceAllocationBoyd} takes up the allocation problem \eqref{eq:resourceallocationproblemIntro} to focus on choosing the proportional weights (to the resource trading) to obtain sufficient conditions for the convergence of the algorithm, and to further improve the rate of convergence. \cite{ResourceAllocationADMMBanjac} considers the dual of the resource allocation problem and derives two methods using the alternating direction method of multipliers (ADMM) algorithm. Also considering the dual problem, \cite{ResourceAllocationUncertainty} includes uncertainties in the individual cost functions and solves the problem using sub-gradient methods on the distributed Lagrangian. 

Mixing economics in the team, \cite{ResourceAllocationCai2012DecentralizedCO} considers a stochastic system in which agents allocate shared system resources in response to customer requests that arrive stochastically over time and introduces the notion of a transfer contract to specify compensation among agents when resources are traded. Each agent has a model of how resources are shared by others and it makes allocation decisions by maximizing its utility function subject to such model. However, it is worth noting that in most of the work on decentralized resource allocation, the amount of resource to be allocated is fixed over the iterations; the agents begin at a feasible solution and move along the resource allocation constraint through the feasible solutions to only minimize the cost function in \eqref{eq:resourceallocationproblemIntro}.

In this work, we approach the resource allocation problem through Centroidal Voronoi Tessellations (CVTs). Even though they date centuries, Voronoi tessellations (VTs) have been found to be immensely helpful in various applications ranging from health to computer graphics to natural sciences. The first documented application of Voronoi tessellations appeared in \cite{JohnSnowCholera} on the 1854 cholera epidemic in London in which it is demonstrated that proximity to a particular well was strongly correlated to deaths due to the disease \cite{LiliJu2011}. In more recent decades, VTs have almost become a common basis tool for path planning algorithms by multi-robot systems in the field of coverage control \cite{CortesMartinezMobileSensing} to such an extent that the VT-based coverage control has been generalized using optimal transport-based control \cite{OptimalTransportCoverageControl}. An adaptive coverage controller is proposed in \cite{AdaptiveConverageControl} where the leader in the leader-follower strategy therein distributes the followers within its obstacle-free sensing range, and the optimized distribution is obtained through a CVT. In their study on optimality of multi-robot coverage control, the authors in \cite{SparsityStructureOptimalityCoverageControl} draw a relationship between CVT configurations and the sufficient condition for optimality through the spatial derivative of the density. 

Despite the wide range of applications of CVTs, to the best of our knowledge, they have not been employed in the resource allocation problem. Our main motivations for employing them in the context of resource allocation problem are minimal communication requirement, robustness, flexibility, scalability and generalizability offered by the CVT framework. To allocate one-dimensional resources, a line graph for the communication network in the team information structure is sufficient to obtain the global optima. We will delve deeper into the advantages and our motivation for using CVT in the resource allocation problem in Section \ref{sec::Preliminaries}. 

To demonstrate our solution to the resource allocation problem using the CVT framework, we consider the application of demand-response in smart grids. We are in the era of a volatile energy market with a looming energy crisis. While the underlying occurrences like geopolitical transitions that cause such crises are beyond local control, the effects are certainly received through the spectrum. In such cases, what one can do locally, on smaller scales is to better employ the resources available at hand. The field of demand-response in smart grids aims to maintain robust operation during such times.

As many parts of the world are gradually moving towards competitive transactive energy markets as a means to generate and procure electricity alongside many of the support services required to operate a power system, many countries are pushing the reform of the electricity power sector very positively. For example, Chile pioneered in the 1980s the deregulation of the electric power industry. In today's U.S. retail electricity market, fourteen states have already adequate retail competition with Texas, Illinois, and Ohio respectively having 100$\%$, 60$\%$, and 50$\%$ of their residential customers receiving service from electricity suppliers \cite{ElecMarketReformChen}. But, even today, most of the customers have very limited \quotes{direct} participation in supporting the grid.

Through the developments in the transactive energy market, there have been some interesting and innovative proposals. \cite{BuiltEnvJournalWILLIAMS2020178} proposes a data-driven method to forecast the electricity demand for decentralized energy management. On the consumption end, although mostly work in progress, peer-to-peer (P2P) electricity trading is gaining momentum with time. Analogous to internet servers and clients, P2P electricity trading is the platform where the end consumer becomes a prosumer (functioning as both energy producer and consumer) and exchanges the remaining electricity with other consumers in the power grid \cite{P2Preview}. A detailed review of existing P2P trading projects is carried out in \cite{p2preview2}. Another major proposal in this direction is load aggregation. As defined in \cite{BEUCaggregators} \quotes{An aggregator is a new type of energy service provider which can increase or moderate the electricity consumption of a group of consumers according to the total electricity demand on the grid. An aggregator can also operate on behalf of a group of consumers producing their own electricity by selling the excess electricity they produce.} A detailed review of the value of aggregators in the electricity market can be found in \cite{aggregatorReview}. 


While such a centralized framework is beneficial in certain applications the cost and the risk of the associated communication overhead can be too high. Such a center-heavy approach also makes the framework vulnerable to attacks due to a single point of failure. The advantages of having communication capabilities between agents reduces such associated risks. One then needs to develop a framework to model the flow of information among the agents and design control laws at global and local levels such that the global and local objectives are achieved. We consider a certain power, generated or negotiated between the aggregator and the utilities, as the resource to be allocated among a team of agents that are building loads, HVACs to be specific. Such thermal loads, due to their latency, inherently allow for further flexibility in the CVT framework.

The paper organization is as follows. We begin with review of some definitions and preliminaries, along with our motivation for employing CVTs to solve this problem in Section \ref{sec::Preliminaries}. Like most work on resource allocation problems, we consider the static allocation problem where the amount of resource to be allocated is fixed, and solve it using a system of non-linear equations in Section \ref{sec::StaticAllocation}. We then move to varying the allocation amount and solve the corresponding dynamic resource allocation problem in Section \ref{sec::DynamicAllocation}. In Section \ref{sec::ApplicationtoDemandResponse} we demonstrate the developed decentralized dynamic resource allocation method on a demand-response problem of power allocation in a group of building loads. Finally, we draw conclusions in Section \ref{sec::conclusions} and present some lines of future work.

\section{Preliminaries}
\label{sec::Preliminaries}

In this Section, we will first review some definitions and background of CVTs in Section \ref{subsec::CVTs}, followed by a brief review on iterative and analytical methods to compute CVTs in Section \ref{subsec::CVTComputation}. Then in Section \ref{subsec:NetworkTheoryBasics}, we define some notations on communication and resource graph employed in this paper.

\subsection{Centroidal Voronoi Tessellations}
\label{subsec::CVTs}

Consider a region $\Omega \in \mathbb{R}^n$ with density $\rho(.)$. Denote a team of $N \in \mathbb{N}$ agents indexed by the set $I_N = \{1,2,\hdots,N\}$. 

\begin{itemize}
    \item[1] Tessellation: $\{V_i\}_{i\in I_N}$ is a tessellation of $\Omega$ if $V_i \cap V_j = \emptyset$ for $i \neq j$, and $\cup_{i\in I} {V}_i = {\Omega}$. 
    \item[2] Voronoi region and generators: The Voronoi region ${V}_{z_i}$ of the Voronoi generator $z_i$ is ${V}_{z_i} = \{x \in \Omega : ||x - z_i|| < ||x-z_j||, \ i \neq j \ \text{and} \ i,j \in I_N \}$. 
    \item[3] Voronoi tessellation: The set of Voronoi regions $\textbf{V}_{\textbf{z}} = \{V_{z_i}\}_{i \in I_N}$ of $\{z_i\}_{i \in I_N}$ is called a Voronoi tessellation $\{ \textbf{z},\textbf{V}_{\textbf{z}}\}$.
\end{itemize}

The mass centroid of a region $V_i \subset \Omega$ under the probability density function $\rho(.)$ is defined as:

\begin{equation}
    z_{V_i,\rho}^c  = \frac{\int_{V_i} x \rho(x) dx}{\int_{V_i}\rho(x) dx}
    \label{eq:MassCentroidDefn}
\end{equation}

\begin{figure}
    \centering
    \includegraphics[width=\columnwidth]{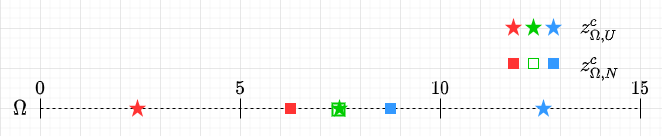}
    \caption{Centroidal Voronoi Tessellations of $[0,15]$ under Uniform and Normal distributions, denoted in star and square symbols respectively.}
    \label{fig:CVT_defn}
\end{figure}

\noindent A Voronoi tessellation in which the generators are the mass centroids of their respective Voronoi regions is called a \textit{Centroidal Voronoi Tessellation} (CVT), \cite{CVT_QiangDu}. The CVT obtained for 3 generators in the region $\Omega = [0,15]$ under Uniform and Normal distributions -- $\mathcal{U}(0,15)$ and $\mathcal{N}(7.5,1)$ -- are shown in Fig. \ref{fig:CVT_defn}. The generators under Uniform ad Normal distribution over $\Omega$ are marked in star and square symbols respectively. 

Consider the functional $\mathcal{F}$ with any $N$ points $\{z_i\}_{i \in I_N} \in \Omega$ and any tessellation $\{V_i\}_{i \in I_N}$ of $\Omega$ as its input arguments:

\begin{equation}
	\mathcal{F}((z_i,V_i), i \in I_N) = \sum_{i \in I_N} \int_{x\in V_i} ||x-z_i||^2 \rho(x) dx
	\label{eq:functionalF_QiangDu}
\end{equation}

\noindent Proposition $3.1$ in \cite{CVT_QiangDu} states that a necessary condition for the function $\mathcal{F}$ to be minimized is that $\{V_i\}_{i \in I_N}$ are the Voronoi regions corresponding to $\{z_i\}_{i \in I_N}$, and simultaneously, $\{z_i\}_{i \in I_N}$ are the centroids of their respective Voronoi regions. In other words, the minimizer of $\mathcal{F}$ is a Centroidal Voronoi Tessellation.

Additionally, if the tessellation in \eqref{eq:functionalF_QiangDu} is fixed to be the Voronoi tessellation of $\{z_i\}_{i \in I_N}$, then the following functional $\mathcal{K}$ has the same minimizer as $\mathcal{F}$, \cite{CVT_QiangDu}.  

\begin{equation}
	\mathcal{K}((z_i), i \in I_N) = \sum_{i \in I_N} \int_{x\in V_{z_i}} ||x-z_i||^2 \rho(x) dx
	\label{eq:functionalK_QiangDu}
\end{equation}

\noindent This functional $\mathcal{K}$ is also referred to as the energy of the tessellation or the quantization energy. 

Including the resource allocation constraint from \eqref{eq:resourceallocationproblemIntro} in the functional $\mathcal{K}$, we obtain the following constrained CVT minimization problem:

\begin{align}
	&\min_{z_i} \sum_{i \in I_N} \int_{x\in V_i} ||x-z_i||^2 \rho(x) dx \nonumber \\
	&\text{s.t,} \nonumber \\
	&\hspace{1cm} \sum_{i \in I_N} z_i = r
	\label{eq:CVTobjectiveSumConstraint}
\end{align}

Comparing with the resource allocation problem \eqref{eq:resourceallocationproblemIntro} we see that the individual objective functions from \eqref{eq:CVTobjectiveSumConstraint} are $\{\int_{x\in V_i} \rho(x) ||x-z_i||^2 dx\}_{i \in I_N}$. While the objective function in \eqref{eq:CVTobjectiveSumConstraint} is separable, a global distribution $\rho(.)$ governs all the individual objective functions. This enables embedding of a desired aggregate behavior in the team through such distributions; the desired aggregate behavior can arise from modeling individual preferences or from an external global trendsetting factor depending on the application at hand.


\subsection{Computation of CVT}
\label{subsec::CVTComputation}

Given $\Omega, N$ and $\rho(.)$, there are various iterative algorithms to compute a CVT in $\Omega$. Under the same conditions, CVT need not be unique for any dimensional region unless certain conditions are imposed on the density function. In 1-D regions, the CVT is unique for log-concave density functions with finite second moment \cite{CVT_Fleischer}. For higher dimensions, finding the conditions on the uniqueness for the general case, without assumptions on the region, density or the number of generators $N$, remains an open area of research. However, it is proved in \cite{UniquenessCVT_Urschel} that for $N=2$, there does not exist a unique CVT for any density for dimensions greater than one.

Accordingly, the solutions rendered by various algorithms to compute the CVT need not be the unique global minimizers, and can converge to a local minima. A deterministic, popular algorithm to obtain a CVT is the Lloyd's algorithm. Introduced in \cite{CVT_Llyod_original} to find the optimal quantization in pulse-code modulation, Lloyd's algorithm has been modified or adapted in various fields. At the core of it, Lloyd's algorithm is an iteration between constructing Voronoi tessellations and their centroids:


\begin{framed}
\noindent Given: $\Omega \subset \mathbb{R}^n$, $N$, $\rho(x)$ \\
Initialize: Generators $\textbf{z} = \{z_i\}_{i \in I}$, where each $z_i \in \Omega$
\begin{itemize}
	\item[1] Construct the Voronoi tessellation $\textbf{V}_\textbf{z}$.
	\item[2] Compute the mass centroids $z^c_{\textbf{V}_{\textbf{z},\rho}}$ of $\textbf{V}_\textbf{z}$. 
	\item[3] If the computed centroids meet certain stopping criteria then terminate. If not, then set $\textbf{z} = z^c_{\textbf{V}_{\textbf{z},\rho}}$, and return to Step 1.
\end{itemize}
 \end{framed}


Even though Lloyd's algorithm is iterative and approximate, it has certain desirable convergence properties. Various global convergence properties of the Lloyd's algorithm are rigorously proved in \cite{ConvergenceLlyod_QiangDu}. Specifically for one-dimensional spaces with log-concave density function, the local convergence using the Lloyd's algorithm has been proved in \cite{CVT_1d_uniquenessKieffer}. Depending on the application at hand, various algorithms that have faster convergence than Lloyd's have been proposed, \cite{CVT_LloydsAlt_FastConvergenceLiu}, \cite{CVT_LlyodsAlt_FastConvergenceWang}, \cite{CVT_LlyodsAlt_FastConvergenceHateley}. Taking the probabilistic approach, \cite{JMacQueen} offers a Monte Carlo sampling based method for the computation of CVTs in any dimension. However, the MacQueen's method only results in the centroids of the CVT and not their Voronoi partitions.

In one-dimensional spaces, we can obtain the entire tessellation analytically using a System of Non-linear Equations (SNLE). The core idea is to parameterize the Voronoi regions in terms of their centroids. In $\Omega = [a,b] \subset \mathbb{R}$, without loss of generality, let the $N$ generators  be $z_1 < z_2 < \hdots < z_N \in \Omega$. Following Section \ref{subsec::CVTs}, by definition, the Voronoi regions are given as: 

\begin{align}
	V_1 &= [a, \frac{z_1+z_2}{2}]  \nonumber \\ 
	&\vdots  \nonumber \\
	V_i &= [ \frac{z_{i-1}+z_i}{2}, \frac{z_i+z_{i+1}}{2}]  \nonumber \\ 
	&\vdots  \nonumber \\
	V_N &= [ \frac{z_{N-1}+z_N}{2}, b]
\label{eq:VoronoiRegionParameterizing_1D}
\end{align}

Additionally, by the definition of CVT the Voronoi generators must be the mass centroids \eqref{eq:MassCentroidDefn}. Rewriting the centroids in terms of the parameterized Voronoi regions from \eqref{eq:VoronoiRegionParameterizing_1D}, we have $\forall i \in I_N$:

\begin{align}
	z_{i}^c  = &\frac{\int_\frac{z_{i-1}^c+z_i^c}{2}^{\frac{z_i^c+z_{i+1}^c}{2}} x \rho(x) dx}{\int_\frac{z_{i-1}^c+z_i^c}{2}^{\frac{z_i^c+z_{i+1}^c}{2}}\rho(x) dx} 
	\label{eq:Centroids_Parameterizing_1D}
\end{align}

where $z_{V_i,\rho}^c$ from \eqref{eq:MassCentroidDefn} is denoted as $z_i^c$ for ease of notation. In \eqref{eq:Centroids_Parameterizing_1D}, there are $N$ number of unknowns: $\{z_i^c\}_{i \in I_N}$, and $N$ equations. Therefore, solving this system of nonlinear equations will result in the centroids of the CVT with which the Voronoi regions can be computed as in \eqref{eq:VoronoiRegionParameterizing_1D}. This is the exact solution of the functional $\mathcal{K}$ from \eqref{eq:functionalK_QiangDu}.

To employ CVTs to solve the resource allocation problem \eqref{eq:resourceallocationproblemIntro}, we treat the Voronoi generators as the agents' resources. In the next Section, we introduce the notations for resource and communication graph for the team which highlights the advantage of employing one-dimensional CVTs in the resource allocation problem.

\subsection{Notations}
\label{subsec:NetworkTheoryBasics}

Let $\textbf{z}=\{z_i\}_{i \in I_N}$ be the agents' resources with each $z_i \in \Omega \subset \mathbb{R}$. Let $\rho(.)$ denote a measure of information or the probability density over $\Omega$. 

Let $\mathcal{Z}$ denote the undirected resource graph and $\mathcal{C}$ denote the undirected communication network of all the agents $i \in I_N$. Their vertex and edge sets are $\{\textbf{z}, \mathcal{E}_Z\}$ and $\{I_N, \mathcal{E}_C\}$, respectively. Denote the set of neighbors of agent $i \in I_N$ according to resource and communication graphs as $\mathcal{N}_{\mathcal{Z}_i}$ and $\mathcal{N}_{\mathcal{C}_i}$, respectively.

The set of resource neighbors  $\mathcal{N}_{\mathcal{Z}_i}$ of each agent $i \in I_N$ is given by \cite{GraphTheoryIntro}:

\begin{align}
    \mathcal{N}_{\mathcal{Z}_i} = \{j \in I_N: z_k < z_j <z_i, \ \forall j,k \in I_N \} \ \ \cup \nonumber \\ \{j \in I_N: z_i < z_j <z_k, \ \forall j,k \in I_N \}  \\
    \implies \ \	j \in \mathcal{N}_{\mathcal{Z}_i} \iff \{z_i,z_j\} \in \mathcal{E}_Z \iff i \in \mathcal{N}_{\mathcal{Z}_j} \nonumber
    \label{eq:resourcenetworkneighbordefinition}
\end{align}

\noindent Parallelly, since $\mathcal{C}$ is undirected, $\{i,j\} \in \mathcal{E}_C \iff i \in \mathcal{N}_{\mathcal{C}_j} $ and $ j \in \mathcal{N}_{\mathcal{C}_i}$, where $\mathcal{N}_{\mathcal{C}_i}$ is the set of communication neighbors of the agent $i$.

Since the resources $z_i \in \mathbb{R}$, the resource network is always a line graph: each agent can have at most $2$ resource neighbors. While the communication network $\mathcal{C}$ can be as complex as a full graph, we set it to be the simplest connected graph in 1-D, $\mathcal{E}_C = \mathcal{E}_Z$. That is, the agents are aware of the resource positions of their neighbors only. 

Any agent being informed of the resource positions of its non-neighbor agent is redundant, since it is not used in the iterative CVT computation methods like Lloyd's. Therefore, if each agent were to communicate only with its resource neighbors, then all the agents would converge to the CVT through Lloyd's algorithm with minimal communication in a decentralized manner. 

In this Section, we looked into how CVTs provide a natural way of embedding a desired distribution in the solution, along with obtaining the solution in a straight-forward decentralized approach with minimal requirements on the team information structure. We studied that one may obtain 1-D CVTs in a decentralized manner using one of the simplest communication graphs: a line graph that is also the same as the resource graph. In the next Section, we take up the resource allocation problem \eqref{eq:resourceallocationproblemIntro} which is a constrained CVT minimization problem, and solve it centrally using the analytical CVT computation method SNLE.

\section{Static resource allocation}
\label{sec::StaticAllocation}

The underlying idea in employing CVTs to solve the resource allocation problem is quite straightforward: the CVT centroids are the resources allocated to the agents, and accordingly, they must sum up to the available amount of resource $r$. Comparing the resource allocation constrained CVT minimization problem \eqref{eq:CVTobjectiveSumConstraint} with \eqref{eq:resourceallocationproblemIntro}, we can observe that the individual agent cost functions are:

\begin{equation}
f_i(z_i) =  \int_{x\in V_i} ||x-z_i||^2 \rho(x) dx 
\label{eq:AgentCostFredholm}
\end{equation}

The main similarity between the constrained CVT minimization problem \eqref{eq:CVTobjectiveSumConstraint} and the resource allocation problem \eqref{eq:resourceallocationproblemIntro} is that the objective functions are separable. That is, the objective functions are decomposed into individual (agent) objective functions that are convex and differentiable. Additionally, the integral equation \eqref{eq:AgentCostFredholm} is known as the Fredholm integral equation of the first kind, \cite{FredholmIntegralEquations}.



Given the separability of \eqref{eq:CVTobjectiveSumConstraint} along with the convexity and differentiability of the agent cost functions, the asymptotic convergence properties developed in \cite{Hariharan} for the resource allocation problem apply to the resource allocation constrained CVT minimization problem \eqref{eq:CVTobjectiveSumConstraint}. The core idea of our solution to the problem of resource allocation through CVTs is to embed the resource allocation constraint within the objective function through the density $\rho(.)$.

Suppose $\rho(.)$ is defined by $n_{\rho}$ number of parameters: $v = (v_1 , v_2 , \hdots, v_{n_{\rho}})$. Let $v_k \in v$ for some $k \in I_{N_{\rho}}$ be an unknown or the \quotes{free} design parameter, and all the other parameters defining the density function be known and fixed. To highlight the dependence of the density function on the free parameter $v_k$, denote the density function as $\rho(x,v_k)$, where $x$ is it's support. 

The optimal solution of the unconstrained CVT minimization problem \eqref{eq:functionalK_QiangDu} is the set of centroids of the Voronoi regions for every agent. Using the definition of centroids \eqref{eq:MassCentroidDefn} for $\{z_i\}_{i \in I_N}$ and embedding the resource allocation constraint transforms the constrained CVT minimization problem into the following system of nonlinear equations with $N+1$ unknowns -- $(z_1^c, z_2^c, \hdots, z_N^c, v_k)$:

\begin{align}
	z_{i}^c  = &\frac{\int_\frac{z_{i-1}^c+z_i^c}{2}^{\frac{z_i^c+z_{i+1}^c}{2}} x \rho(x,v_k) dx}{\int_\frac{z_{i-1}^c+z_i^c}{2}^{\frac{z_i^c+z_{i+1}^c}{2}}\rho(x,v_k) dx} \ \ \ \forall i \in I_N   \nonumber \\
	&\sum_{i=1}^N  z_i^c = r
	\label{eq:Nplus1SNLE_summationsconstraint}
\end{align}

The solution of this system of nonlinear equations, $(z_1^c, z_2^c, \hdots, z_N^c, v_k)$, satisfies the following:
\begin{itemize}
	\item $(z_1^c, z_2^c, \hdots, z_N^c)$ are the $N$ centroids of the CVT in $\Omega = [a,b]$ under the density function $\rho(x,v_k)$.
	\item The centroids sum up to $r$, satisfying the resource allocation constraint in \eqref{eq:resourceallocationproblemIntro}.
\end{itemize}

The main solution of interest here is the solved design parameter $v_k$ which is fed to the Lloyd's algorithm in its initialization step. In that case, all the agents can still maintain communication only with their resource neighbors, and since they are all initialized with the same design parameters, all the agents are minimizing the cost function \eqref{eq:functionalK_QiangDu} under the same specifications, and obtain the CVT.

\begin{figure}[h]
	\includegraphics[width=\columnwidth]{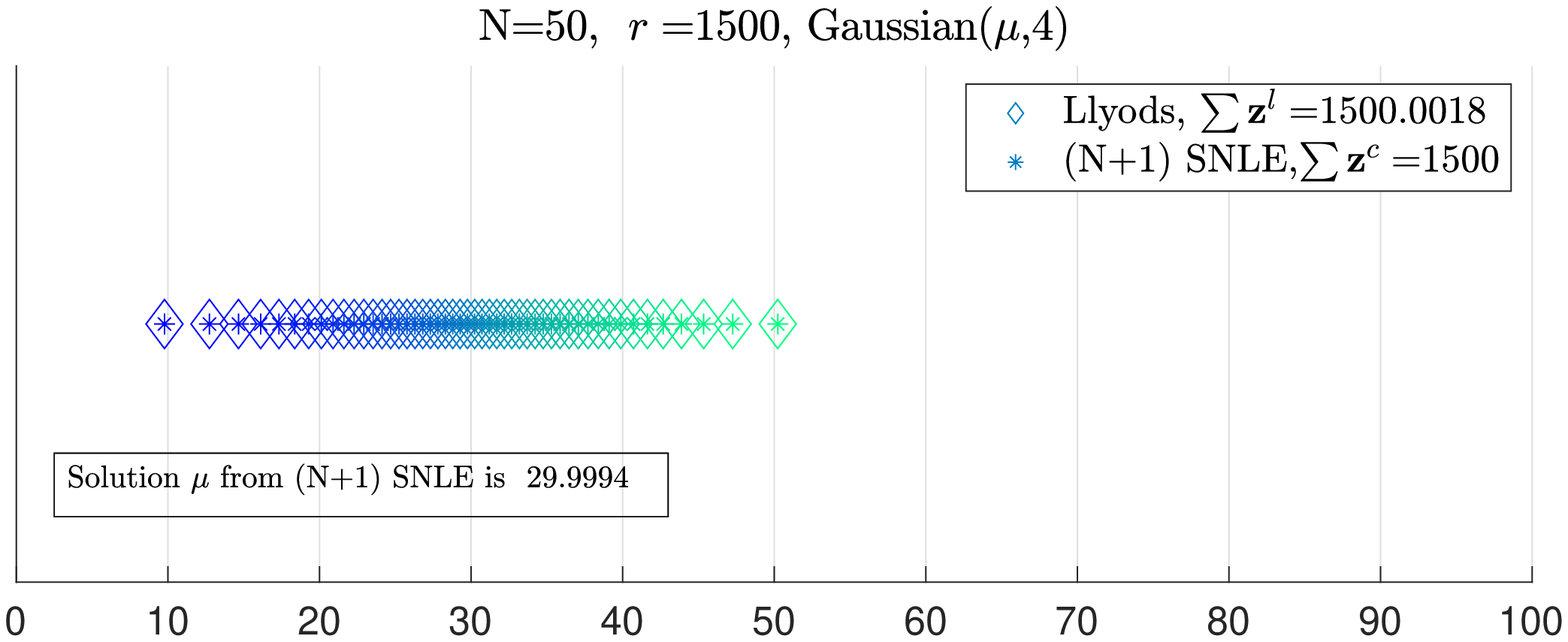}
	\includegraphics[width=\columnwidth]{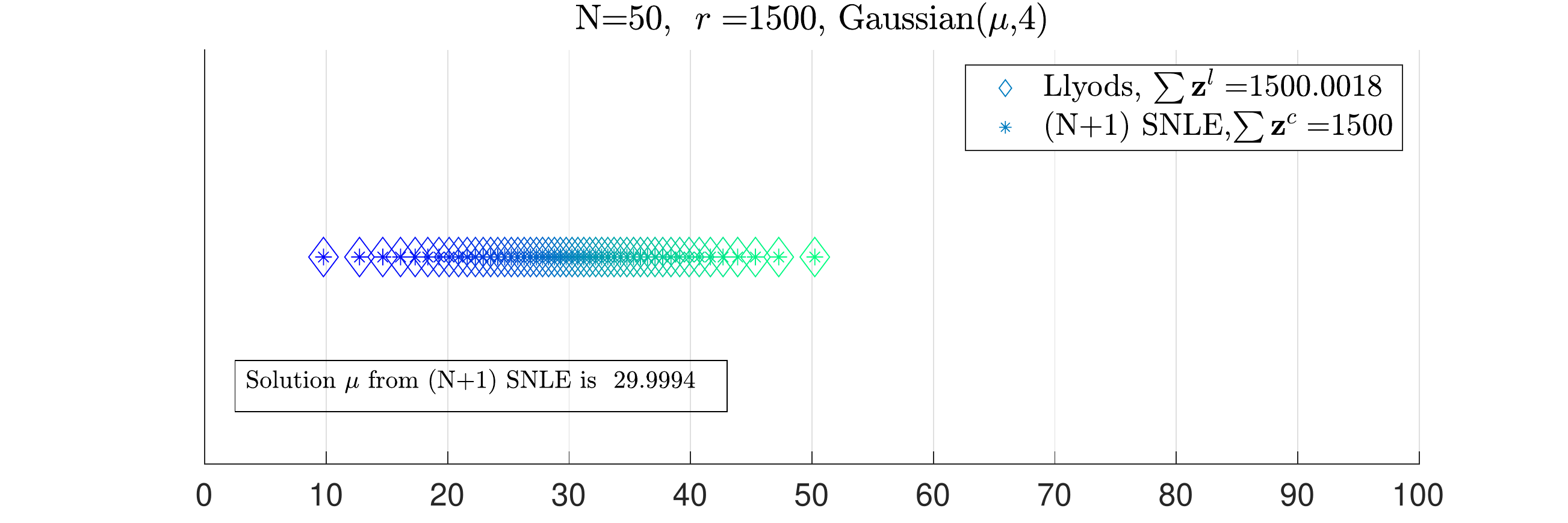}
	\includegraphics[width=\columnwidth]{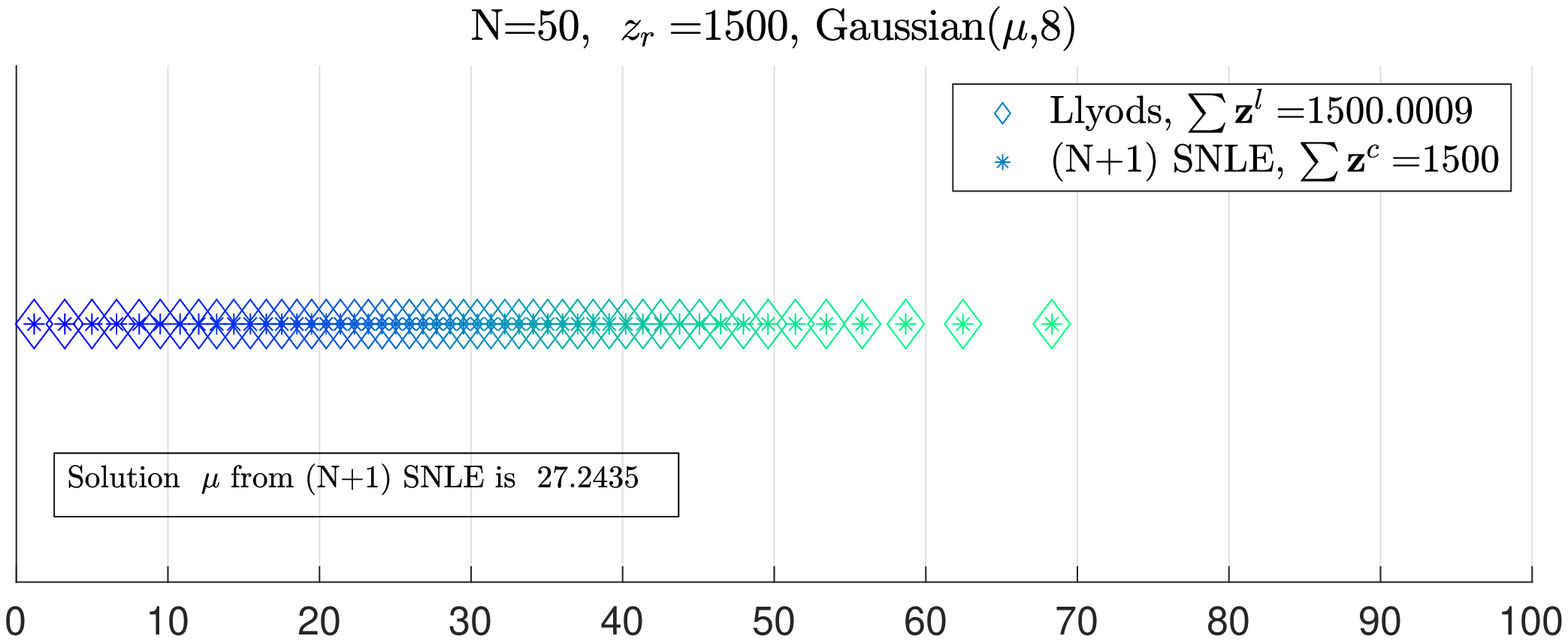} 
	\caption{Allocation of $r$ amount of resource among 50 agents in $\Omega = [0,100]$ under Gaussian distribution for specified variances -- $4$ (top and middle) and $8$ (bottom). The mean of the distribution $\mu$ is the solution $v_k$ from \eqref{eq:Nplus1SNLE_summationsconstraint}.}
	\label{fig:N50_Gaussian_O0100_Np1}
\end{figure}

We now demonstrate the method with different simulation cases. In Fig. \ref{fig:N50_Gaussian_O0100_Np1}, the region $\Omega = [0,100], \ N=50$, and the density is Gaussian. Out of the three examples therein, the top two have the same variance but are required to allocate different amounts of resources -- $2500$ in the first and $1500$ in the second -- among the same number of agents.  Accordingly, we can observe the resources allocated among all the agents are lower in the second case than the first. Moving from the second example to the third (the bottom graph in Fig. \ref{fig:N50_Gaussian_O0100_Np1}), the variance is increased while keeping all other parameters the same. In all these three cases, the free design parameter $v_k$ is $\mu$ -- the mean of the Gaussian distribution. The solution of the free parameter obtained from solving the $N+1$ equations from \eqref{eq:Nplus1SNLE_summationsconstraint}, is shown in the figures and is used to initialize the Lloyd's algorithm. The generators obtained from the Lloyd's algorithm and the generators from solving \eqref{eq:Nplus1SNLE_summationsconstraint} are plotted together. We can observe that the two solutions are very close to each other. Additionally, both the solutions sum up to the resource to be allocated -- $r$, with an acceptable error. 

\begin{figure}[h]
	\includegraphics[width=\columnwidth]{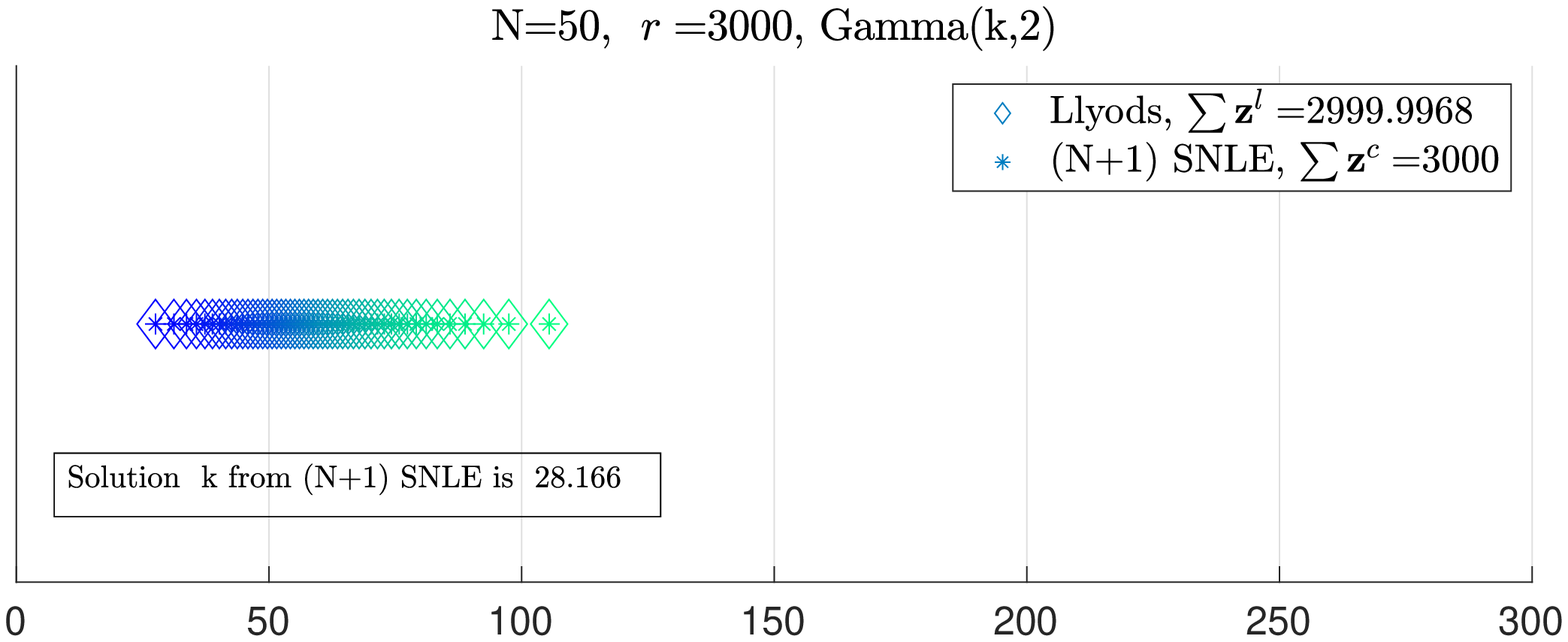}
	\includegraphics[width=\columnwidth]{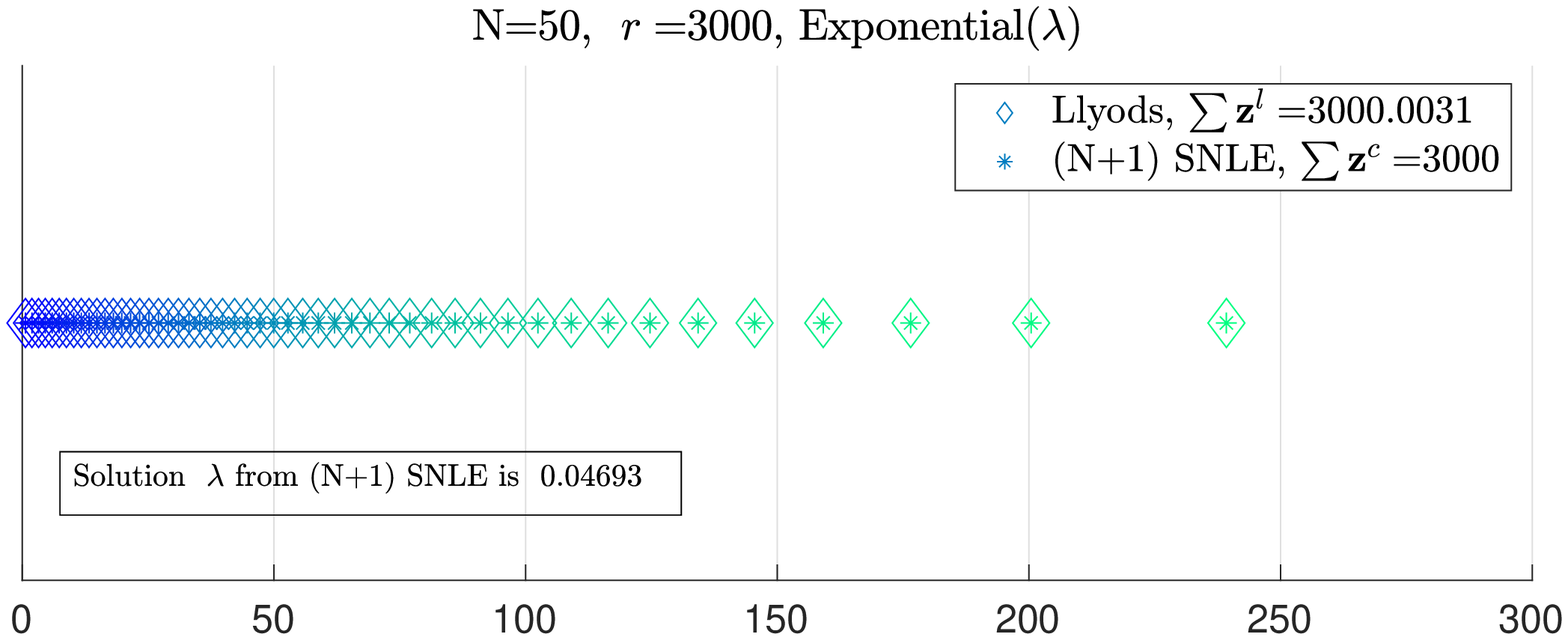}
	\includegraphics[width=\columnwidth]{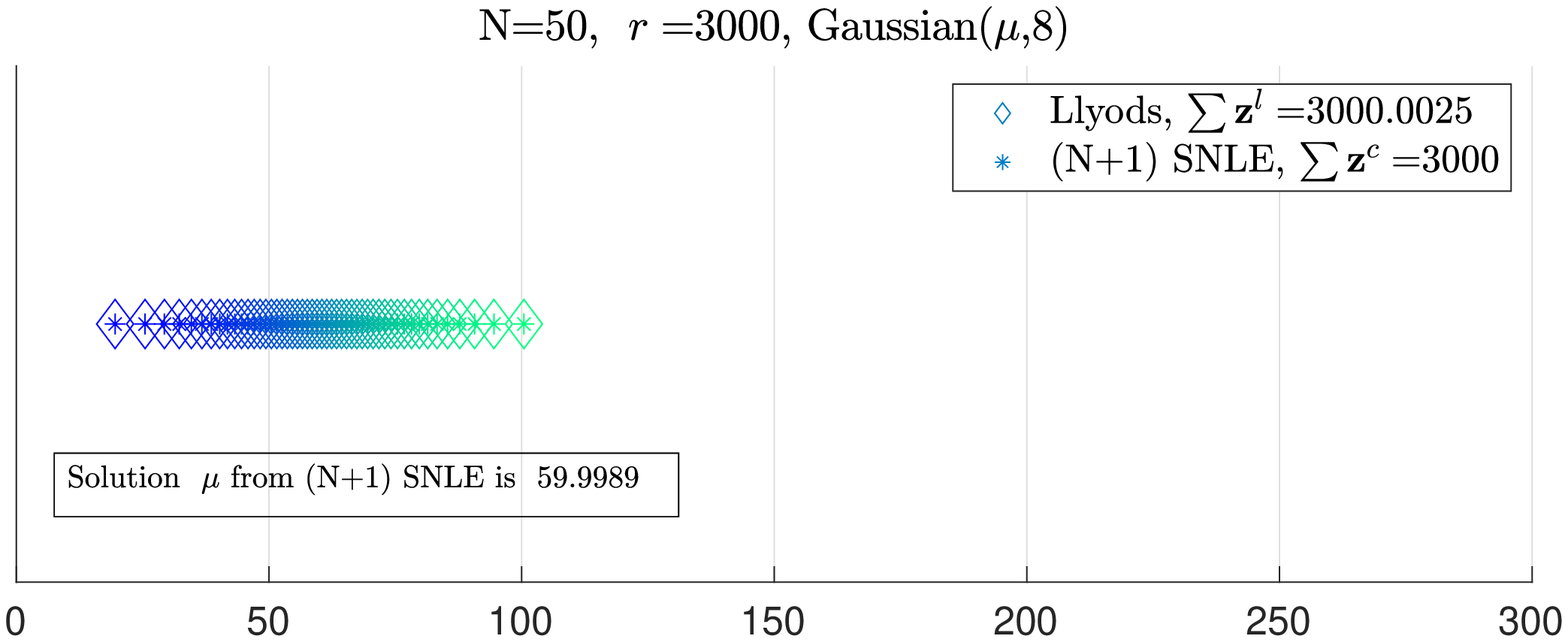}
	\caption{Allocation of $r$ amount of resource among 50 agents in $\Omega = [0,300]$ under three different distributions. Top: Gamma distribution with the free parameter $v_k$ being $k$. Middle: Exponential distribution with the free parameter $v_k$ being $\lambda$. Bottom: Gaussian distribution with the free parameter $v_k$ being $\mu$.}
	\label{fig:N50__O0300_Np1}
\end{figure}

Similarly, we present another set of simulations in Fig. \ref{fig:N50__O0300_Np1}. In the three cases therein, $\Omega, N$ and $r$ are the same. The difference in the three cases is the underlying distributions -- Gamma distribution in the top figure, Exponential in the middle, and Gaussian distribution in the bottom figure. Like in Fig. \ref{fig:N50_Gaussian_O0100_Np1}, the solutions from the two approaches are close to each other and also sum up to $r$. 

Even though in this approach we obtain the solution of \eqref{eq:CVTobjectiveSumConstraint}, the SNLE method is centralized and its grows with $N$. However, it is worth pointing out that once the Lloyd's algorithm is initialized with a design parameter $v_k$, CVT obtained using Lloyd's algorithm is scalable to any $N$ because regardless of the total number of agents $N$, each agent can have at most two neighbors. 

Like most of the solutions to the resource allocation problem, in this Section we considered a fixed amount of resource to be allocated among al the agents. Using the developed static allocation method as the initialization step, in the next Section we consider the problem of dynamic resource allocation problem where the amount of resource to be allocated is time-varying and all the agents are aware of the quantity.


\section{Dynamic Resource Allocation}
\label{sec::DynamicAllocation}

In the previous section, we solved the static resource allocation problem by using the centralized system of nonlinear equations. However, extending the same approach to varying amount of resource-to-be-allocated results in a centralized approach. Therefore, in this Section we focus on developing a decentralized approach to the dynamic resource allocation problem.

Employing the static resource allocation problem as the initialization step, our solution approach to the dynamic resource allocation problem under Normal distribution involves a one-step update that maintains the dynamic resource allocation constraint while preserving the CVT. We employ the following Lemma \ref{lemma_DynamicAllocationNormal} to obtain such one-step update in Theorem \ref{theorem_DynamicAllocationNormal}. Through the design process, we assume that the amount of resource to be allocated among all the agents over the considered time duration is known to all the agents. 

Suppose $\rho(.) = \mathcal{N}(\mu,\sigma^2)$ and the \quotes{free} parameter $v_k$ is $\mu$. Then we have:

\begin{lemma}
Suppose at time $k$, $\{z_i(k)\}_{i \in I_N}$ are the centroids of the CVT in $\Omega \subset \mathbb{R}$ with density $\rho(.) = \mathcal{N}(\mu(k),\sigma^2)$. Then the following relationship holds between the time-updated centroids:

\label{lemma_DynamicAllocationNormal}
\end{lemma}

\begin{align}
z_i(k+1) - z_i(k) &= z_j(k+1) - z_j(k) \nonumber \\
&= \mu(k+1) - \mu(k) = -\delta 
\label{eq:ziupdatedifference_muchange_LemmaStatement}
\end{align}

\textbf{Proof:} Let $\mu(k+1) = \mu(k) -\delta$. Because $\{z_i(k)\}_{i \in I_N}$ are the centroids with normal distribution, we have by definition:

\begin{align}
z_i(k) = \frac{\int_{V_i(k)} x e^{\frac{(x-\mu(k))^2}{2\sigma^2}} dx}{\int_{V_i(k)} e^{\frac{(x-\mu(k))^2}{2\sigma^2}} dx} \nonumber 
\end{align}

Similarly, writing out the mass centroid for the next time instant $k+1$ using $\mu(k+1) = \mu(k) -\delta$, we have:

\begin{equation}
z_i(k+1) = \frac{\int_{V_i(k+1)} x e^{\frac{(x-(\mu(k)-\delta))^2}{2\sigma^2}} dx}{\int_{V_i(k)} e^{\frac{(x-(\mu(k)-\delta))^2}{2\sigma^2}} dx} 
\end{equation}

Suppose $V_i(k) = [a,b] \subset \Omega$. Consider the change of variables $y = x - \delta$. Then the mass centroids transform as:

\begin{align}
z_i(k) &= \frac{\int_{a}^b x e^{\frac{(x-\mu(k))^2}{2\sigma^2}}  dx}{\int_{a}^b e^{\frac{(x-\mu(k))^2}{2\sigma^2}}  dx} \nonumber \\
&= \frac{\int_{a-\delta}^{b-\delta} (y+\delta) e^{\frac{(y+\delta-\mu(k))^2}{2\sigma^2}} dy}{\int_{a-\delta}^{b-\delta} e^{\frac{(y+\delta-\mu(k))^2}{2\sigma^2}} dy} \nonumber \\
&= \frac{\int_{a-\delta}^{b-\delta} (y+\delta) e^{\frac{(y-(\mu(k)-\delta))^2}{2\sigma^2}} dy}{\int_{a-\delta}^{b-\delta} e^{\frac{(y-(\mu(k)-\delta))^2}{2\sigma^2}} dy} \nonumber \\
&= \frac{\int_{a-\delta}^{b-\delta} y e^{\frac{(y-(\mu(k)-\delta))^2}{2\sigma^2}} dy + \int_{a-\delta}^{b-\delta} \delta e^{\frac{(y-(\mu(k)-\delta))^2}{2\sigma^2}} dy}{\int_{a-\delta}^{b-\delta} e^{\frac{(y-(\mu(k)-\delta))^2}{2\sigma^2}} dy} \nonumber \\
 &= \frac{\int_{V_i(k+1)} y e^{\frac{(y-\mu(k+1))^2}{2\sigma^2}} dy}{\int_{V_i(k+1)} e^{\frac{(y-\mu(k+1))^2}{2\sigma^2}} dy} + \delta \frac{\int_{V_i(k+1)} e^{\frac{(y-\mu(k+1))^2}{2\sigma^2}} dy}{\int_{V_i(k+1)} e^{\frac{(y-\mu(k+1))^2}{2\sigma^2}} dy} \nonumber \\
 &= z_i(k+1) + \delta \nonumber \\
  \implies &z_i(k+1) - z_i(k) = -\delta 
  \label{eq:ziupdatedifference_muchange}
\end{align}

Since \eqref{eq:ziupdatedifference_muchange} holds for all $i \in I_N$ and $\mu(k+1) = \mu(k) -\delta$, we have \eqref{eq:ziupdatedifference_muchange_LemmaStatement} proved.

$\hfill \square$

\vspace{0.1cm}

\begin{theorem}
Suppose we are initialized with static resource allocation solution at discrete-time $k$ and are at the following conditions: $\{z_i(k)\}_{i \in I_N} \ s.t \ \sum_{i \in I_N} z_i(k) = r(k), \ \ \{z_i(k)\}_{i \in I_N} \sim \mathcal{N}(\mu(k),\sigma^2)$. Suppose the resource to be allocated at the next time instant is $r(k+1)$. If agents update their resources as

\begin{equation}
z_i(k+1) = z_i(k) + \frac{1}{N}(r(k+1) - r(k))
\label{eq:AgentUpdateDynamicAllocation}
\end{equation}

then the resulting solution satisfies the following:
\begin{itemize}
\item [1] $\sum_{i \in I_N} z_i(k+1) = r(k+1)$
\item [2] $\{z_i(k+1)\}_{i \in I_N} \sim \mathcal{N}(\mu(k+1),\sigma^2)$
\end{itemize}

\noindent where $\mu(k+1)$ is a solution of the $(N+1)$ SNLE \eqref{eq:Nplus1SNLE_summationsconstraint}.

\label{theorem_DynamicAllocationNormal}
\end{theorem}

\textbf{Proof:}
Obtain the time-difference of the summation of the resources:

\begin{align}
\sum_{i \in I_N} z_i(k+1) - \sum_{i \in I_N} z_i(k) 
 &= \sum_{i \in I_N} z_i(k+1) - z_i(k) \nonumber \\
 &= - N \delta \nonumber \\
 &= N (\mu(k+1) - \mu(k)) \nonumber \\
 &= r(k+1) - r(k) 
\end{align}

$\hfill \square$

\vspace{0.1cm}

Following Theorem \ref{theorem_DynamicAllocationNormal} we obtain the CVT that satisfies the dynamic resource allocation constraint  for the desired Normal distribution in a decentralized manner. While this fulfills our objective, it can be observed that the approach is quite rigid. In practical applications where the agents have their own set of dynamics and are trying to navigate around certain local objectives as well, this approach can be restrictive. Therefore, to extend its applicability we introduce flexibility in the design by allowing for (local) negotiations between neighbors through what we call a \quotes{\textit{civility model}}. 

Before detailing the civility model, let us introduce some new notations. For each agent $i \in I_N$, denote its desired resource amount at time $k$ that meets its local objective as $u_i(k)$. For example, if the agent $i$ is responsible for the control of a certain system modeled as a state-space, such $u_i(k)$ could be the control input from a state-feedback controller, from an LQR or from any such local controller. Since we are operating in 1-D spaces, recall from Section \ref{subsec:NetworkTheoryBasics} that the resource and communication graphs are the same. Following the same notation therein, denote the communication graph at time $k$ as $\mathcal{C}^k$, and the neighbors of agent $i$ at time $k$ as $\mathcal{N}_{\mathcal{C}^k_i}$. 

\vspace{0.3cm}

\noindent \textbf{Initialization}: All agents are aware of the total resources $r(k), \forall k \in T$ and the initial communication network $\mathcal{C}^{k-1}$. Solve the static allocation problem for the resource $r(k-1)$. 

Following the initialization, the civility model for local negotiations is developed as follows.

\begin{framed}
\hspace{0.5cm} \textbf{Civility model for local negotiations}

\vspace{0.1cm}
\noindent For every agent $i \in I_N$, at every time $k \in T$, do:
\begin{itemize}
\item[1] Compute the resource update $z_i(k)$ from \eqref{eq:AgentUpdateDynamicAllocation}. Compute $u_i(k)$ based on the local requirements, possibly from the local controller.
\item[2] Compute the neighbor of interest as $\hat{j} = \{j \in \mathcal{N}_{\mathcal{C}^k_i} \cup i \text{ such that }  ||u_i(k) - z_{\hat{j}}(k)|| < ||u_i(k) - z_j(k)|| \}$.
\item[3] Swap resources with the neighbor of interest $\hat{j}$ from the previous step, if $\hat{j}$ indicates it has not already been taken. This results in $z_i(k) = z_{\hat{j}}(k)$. If $\hat{j}$ has already negotiated with its other neighbor and is hence taken, or if $\hat{j} = i$, then implement the resource update $z_i(k)$ from Step 1.
\end{itemize}
\end{framed}

\noindent It is worth noting that the communication network is dynamically updated in a decentralized manner, and that such an update naturally follows from the resource swap during the local negotiations. We call this approach the civility model because if a neighbor asks to swap, the agent complies with it regardless of its own local requirement. And hence, since all the agents follow the same model, no agent is at a disadvantage in following such an approach. 

To demonstrate the clarity and effectiveness of the proposed method to dynamically allocate resources in a decentralized manner, we consider the application of demand-response in smart grids. Specifically, we consider a group of Heating, Ventilation, and Air Conditioning (HVAC) units that have their local objectives to maintain their indoor air temperatures according to certain desired setpoints, but are also required to respond to certain demand (power) curve by consuming the available power as a team of agents.

\section{Application to Demand Response}
\label{sec::ApplicationtoDemandResponse}

To demonstrate the developed method, we consider power allocation in a group of building HVACs. In this application of demand-response, the agents are the building HVACs. The resources to be allocated to all the agents are the powers consumed by the HVACs to maintain the local indoor air temperatures. We adapt the state-space model from \cite{XiaoMa_HVACmodel} to simulate the indoor air temperatures for each agent $i$ as:

\begin{align}
\dot{x}_i(t) = A_i x_i(t) + B_i u_i(t) + G_i w_i(t) \nonumber \\
y_i(t) = C_i x_i(t) + D_i u_i(t)
\label{eq:HVACmodel}
\end{align}

\noindent The input $u_i$ is the power consumption of the HVAC (agent $i$), the output $y_i$ is the indoor air temperature, and $w_i$ is the vector of disturbances -- outdoor air temperature and solar radiation. The system matrices for each agent are given by:

\begin{equation*}
A_i = 
\begin{bmatrix}
\frac{-(K_{1}^i+K_{2}^i+K_{3}^i+K_{5}^i) }{C_{1}^i} & \frac{(K_{1}^i+K_{2}^i) }{C_{1}^i}  & \frac{K_{5}^i}{C_{1}^i} \\
\frac{K_{1}^i+K_{2}^i}{C_{2}^i}   & \frac{-(K_{1}^i+K_{2}^i)}{C_{2}^i}     & 0 \\
\frac{K_{1}^i}{C_3^i}    &  0    &  \frac{-(K_{4}^i+K_{5}^i)}{C_{3}^i}
\end{bmatrix}
\end{equation*}

\begin{align*}
B_i = 
\begin{bmatrix}
\frac{1}{C_1^i}+\frac{1}{C_2^i}\\
0 \\
0
\end{bmatrix}
G_i = 
\begin{bmatrix}
\frac{K_3^i}{C_1^i}  & \frac{1}{C_1^i}  \\
0 &  \frac{1}{C_2^i} \\
\frac{K_4^i}{C_3^i}   & 0 
\end{bmatrix} 
C_i = 
\begin{bmatrix}
1 &  0  & 0
\end{bmatrix} \\
\end{align*}

\noindent with $D_i$ being a zero matrix. The system parameters, which are resistances and capacitances in the thermal dynamics of the building model, for each agent $i$ are obtained as realizations of the following normal distributions:

\begin{align*}
&K_1  \sim \mathcal{N}(16.48,0.1) \hfill &K_5  \sim \mathcal{N}(23.04,0.1) \\
&K_2  \sim \mathcal{N}(108.5,0.1) \hfill &C_1  \sim \mathcal{N}(9.36 \times 10^5,1) \\
&K_3  \sim \mathcal{N}(5,0.1) \hfill &C_2  \sim \mathcal{N}(2.97 \times 10^6,1) \\
&K_4  \sim \mathcal{N}(30.5,0.1) \hfill &C_3  \sim \mathcal{N}(6.695 \times 10^5,1)
\end{align*}

\begin{figure}[h]
\centering
\includegraphics[width=\columnwidth]{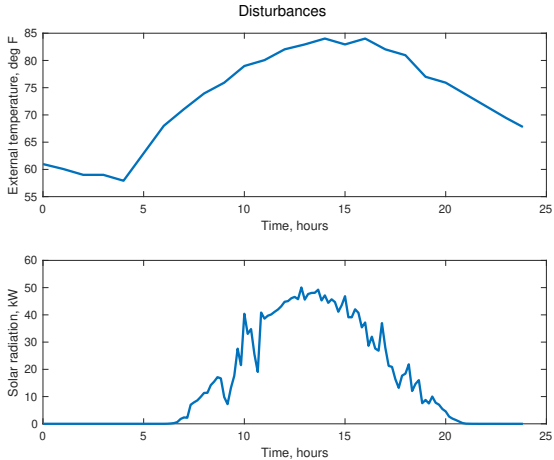}
    \caption{Disturbances in the HVAC model \eqref{eq:HVACmodel}}
    \label{fig:DR_disturbances}
\end{figure}

We implement the agent's model by discretizing the state-space model \eqref{eq:HVACmodel} with a sampling time of 10 minutes. In the HVAC model, the input $u_i$ corresponds to cooling when negative and to heating when positive. Regardless, its absolute value is the power consumed, and therefore we use that for local negotiations and let the individual agent decide whether to use the allocated power for heating or cooling based on its local control. To maintain the indoor air temperatures from a local control, we employ a state-feedback controller for pole placement for every agent to determine its $u_i(k)$. We consider the same disturbances for all the agents; the outdoor air temperature and the solar radiation, \cite{KadirMFCpaper}, we use for our simulations are shown in Fig. \ref{fig:DR_disturbances}.

To begin the dynamic resource allocation we initialize $\rho(.)$ as $\mathcal{N}(\mu,\sigma^2)$, and following Section \ref{sec::StaticAllocation}, solve the first-time allocation (initialization) as a static allocation problem. Communicating to all the agents the resulting mean $\mu$, we begin the decentralized dynamic allocation as laid out in Section \ref{sec::DynamicAllocation}. 

\begin{figure}[h]
    \centering
    \includegraphics[width=\columnwidth]{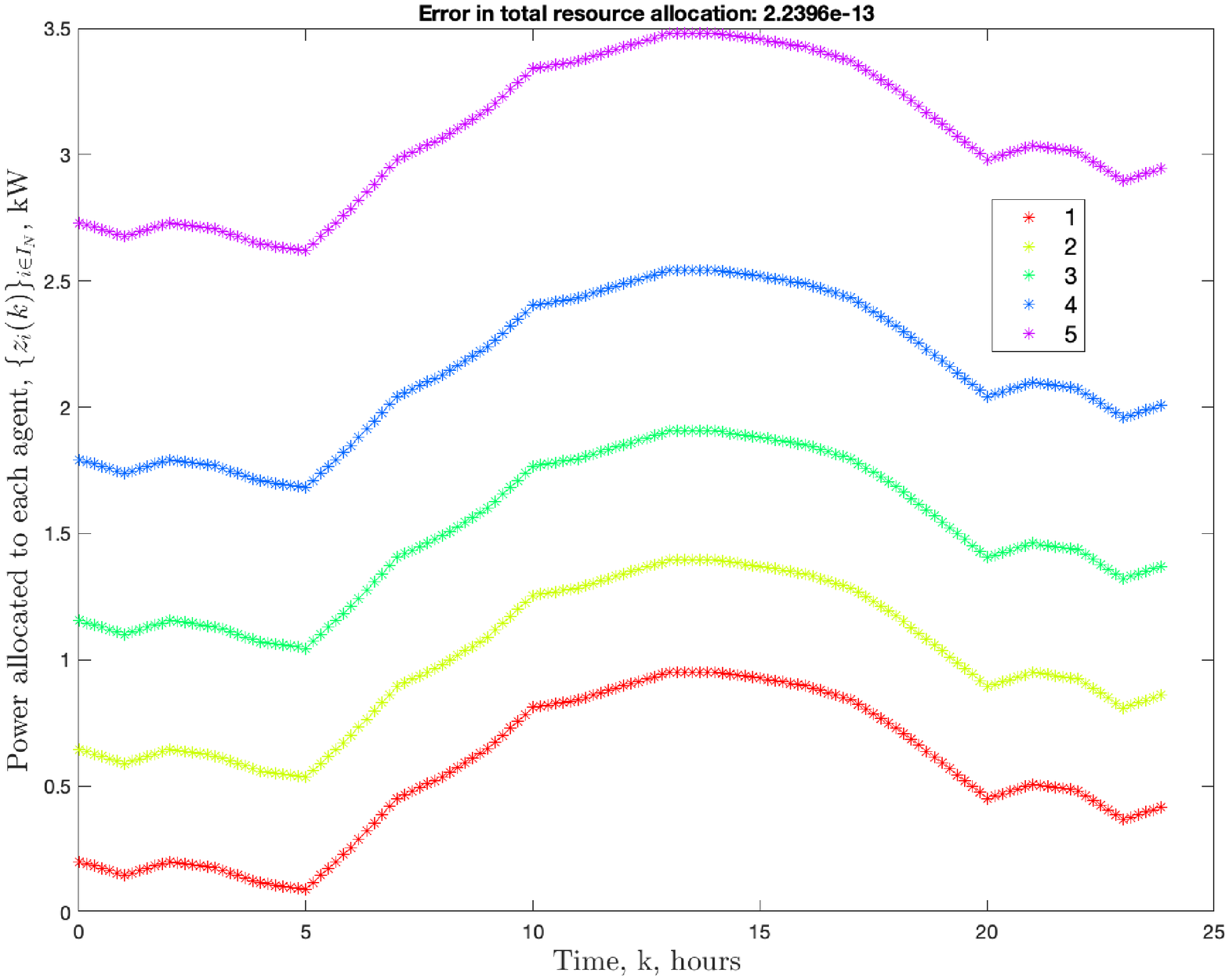}
    \includegraphics[width=\columnwidth]{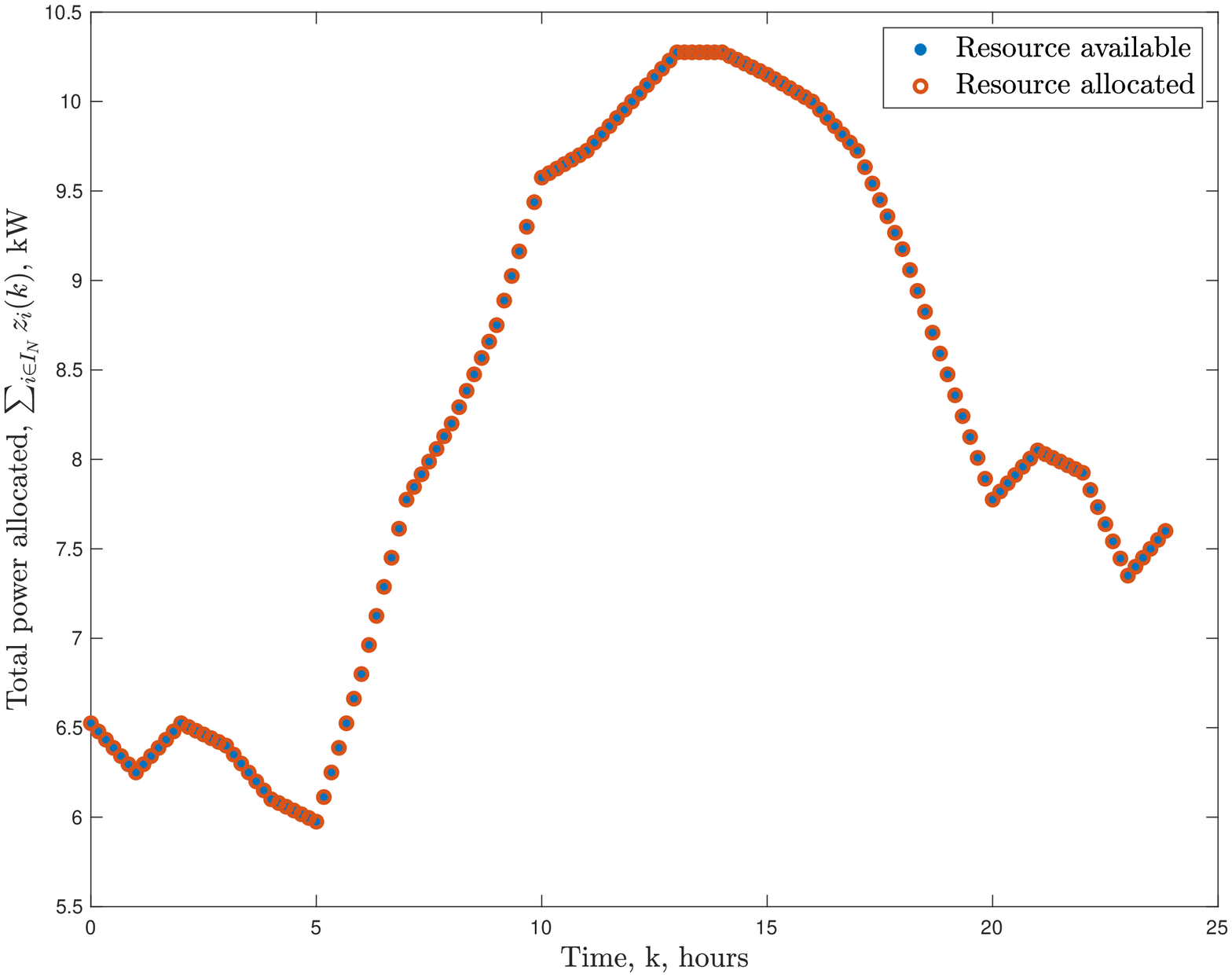}
    \caption{Baseline power consumptions: Agents acting based on the resource allocation constraint without the civility model. Left: Individual power consumption. Right: Total power consumption.}
    \label{fig:DynamicAllocation_N5Gaussian_LocalCtr_NoSwaps_Powers}
\end{figure}

\begin{figure}
\centering
 \includegraphics[width=\columnwidth]{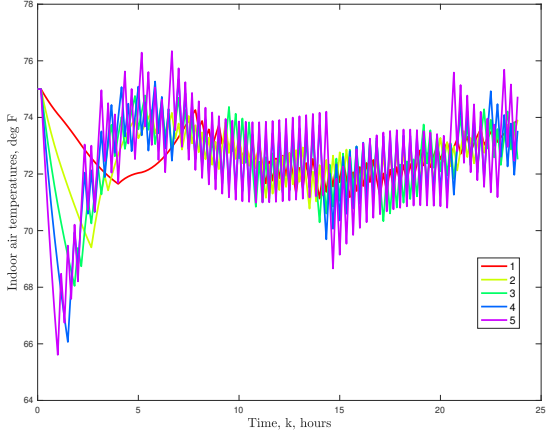}
    \caption{Baseline indoor air temperatures: Agents acting based on the resource allocation constraint without the civility model.}
    \label{fig:DynamicAllocation_N5Gaussian_LocalCtr_NoSwaps_Temps}
\end{figure}

Even though the performance of the developed approach depends on the total available resource and the local requirements, the civility model allows for flexibility, and that could be necessary to compensate for local disturbances or for improper selection of the (desired) distribution in the tessellation. To explain the graphical setup of our results, we begin with $N=5$ in Fig \ref{fig:DynamicAllocation_N5Gaussian_LocalCtr_NoSwaps_Powers}. The top figure shows the power consumption of all the agents at every time instant, and the bottom figure shows their total power consumption versus the available power. Augmenting, Fig \ref{fig:DynamicAllocation_N5Gaussian_LocalCtr_NoSwaps_Temps} shows the individual indoor air temperatures when the agents implement the allocated power from Fig \ref{fig:DynamicAllocation_N5Gaussian_LocalCtr_NoSwaps_Powers}. 

\begin{figure}
    \centering
    \includegraphics[width=\columnwidth]{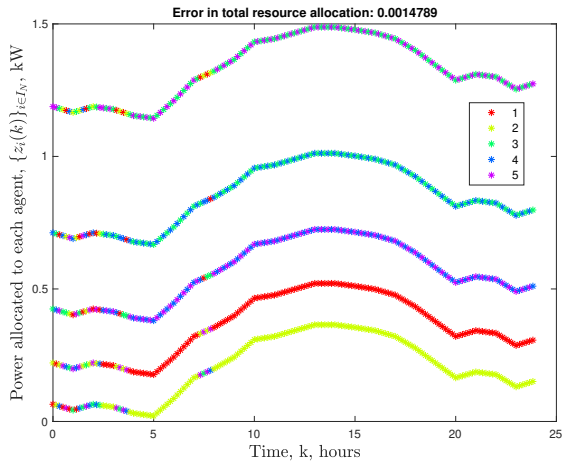}
    \includegraphics[width=\columnwidth]{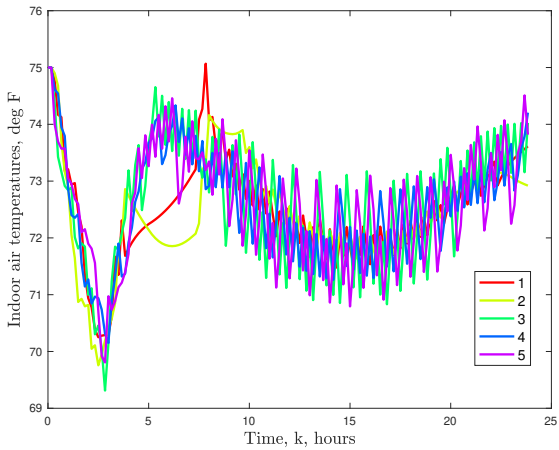}
    \caption{Civility model with local state feedback controller for $5$ agents.}
    \label{fig:DynamicAllocation_N5Gaussian_LocalCtr_Swaps}
\end{figure}

Next, we demonstrate the civility model from Section \ref{sec::DynamicAllocation} by allowing for swapping through local negotiations. Continuing the previous case we first consider only $5$ agents in the team in Fig. \ref{fig:DynamicAllocation_N5Gaussian_LocalCtr_Swaps} and then demonstrate for $15$ agents. For every agent, the power consumptions and the indoor air temperatures are shown in the same color throughout the simulation duration. For example, agent $2$ is shown in red. Thus one can follow the agents' negotiations and the resulting swaps and communication network by following the individual power consumption of the agents through their colors. In the subsequent cases, we do not show the satisfaction of the resource allocation constraint through a dedicated figure since we can concisely express it numerically as the error between total power consumption of all the agents and the available power; we use $l_2$ norm to compute the power consumption error.

\begin{figure}
    \centering
    \includegraphics[width=\columnwidth]{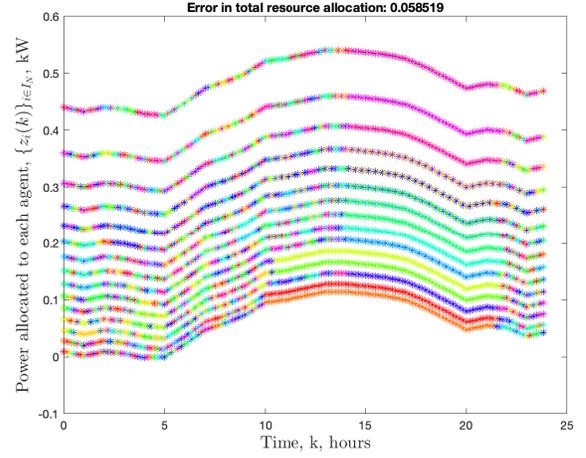}
    \includegraphics[width=\columnwidth]{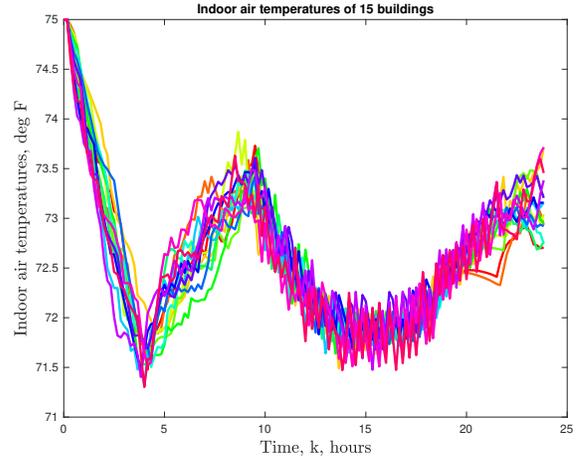}
    \caption{Civility model with local state feedback controller for $15$ agents. The temperature setpoints for all the HVACs are at $72\degree F$.}
    \label{fig:DynamicAllocation_N15Gaussian_LocalCtr_Swaps}
\end{figure}

The strengths of the developed method lie in its robustness in maintaining the resource allocation constraint while accounting for local preferences in a truly decentralized manner. To demonstrate the same, we perturb the setpoints of certain agents and observe the corresponding resource negotiations and the air temperatures in Fig \ref{fig:DynamicAllocation_N15Gaussian_LocalCtr_Swaps}. We can observe the increased amount of negotiations in the increased number of swaps spreading throughout the team to correct for the disturbances for some of the agents. Quantifying the swaps, we have that out of $144$ time-steps in the simulation, each agent swapped $126.9$ times on average and that every agent has been a neighbor of almost every other agent. This suggests a high degree of variation in the communication network, further suggesting that the amount of information is so fragmented among all the agents that it is sufficient to meet the resource allocation constraint while following the desired distribution in the tessellation but not enough for any agent to recreate the behavior of any other agent.

\begin{figure}
    \centering
    \includegraphics[width=\columnwidth]{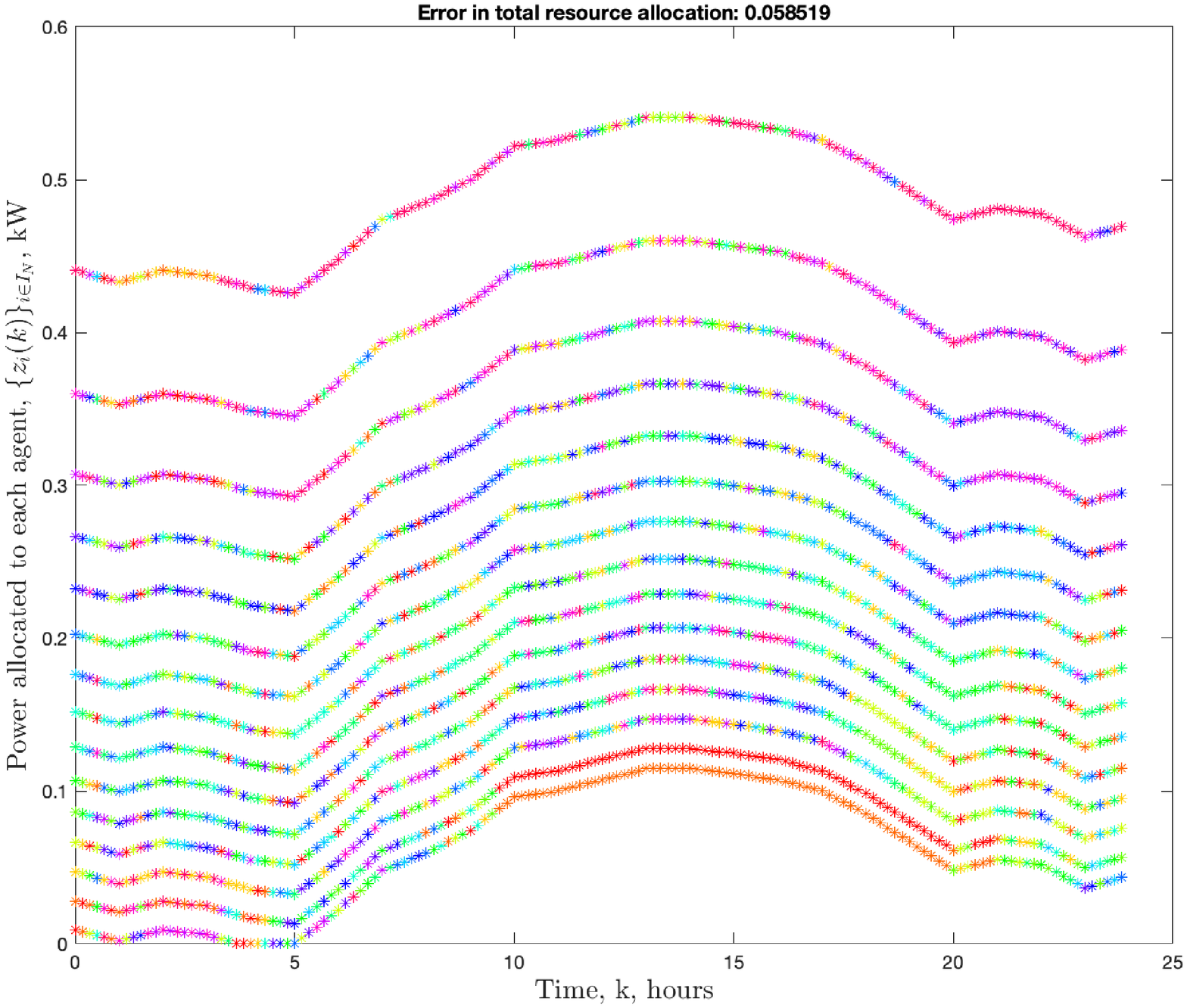}
    \includegraphics[width=\columnwidth]{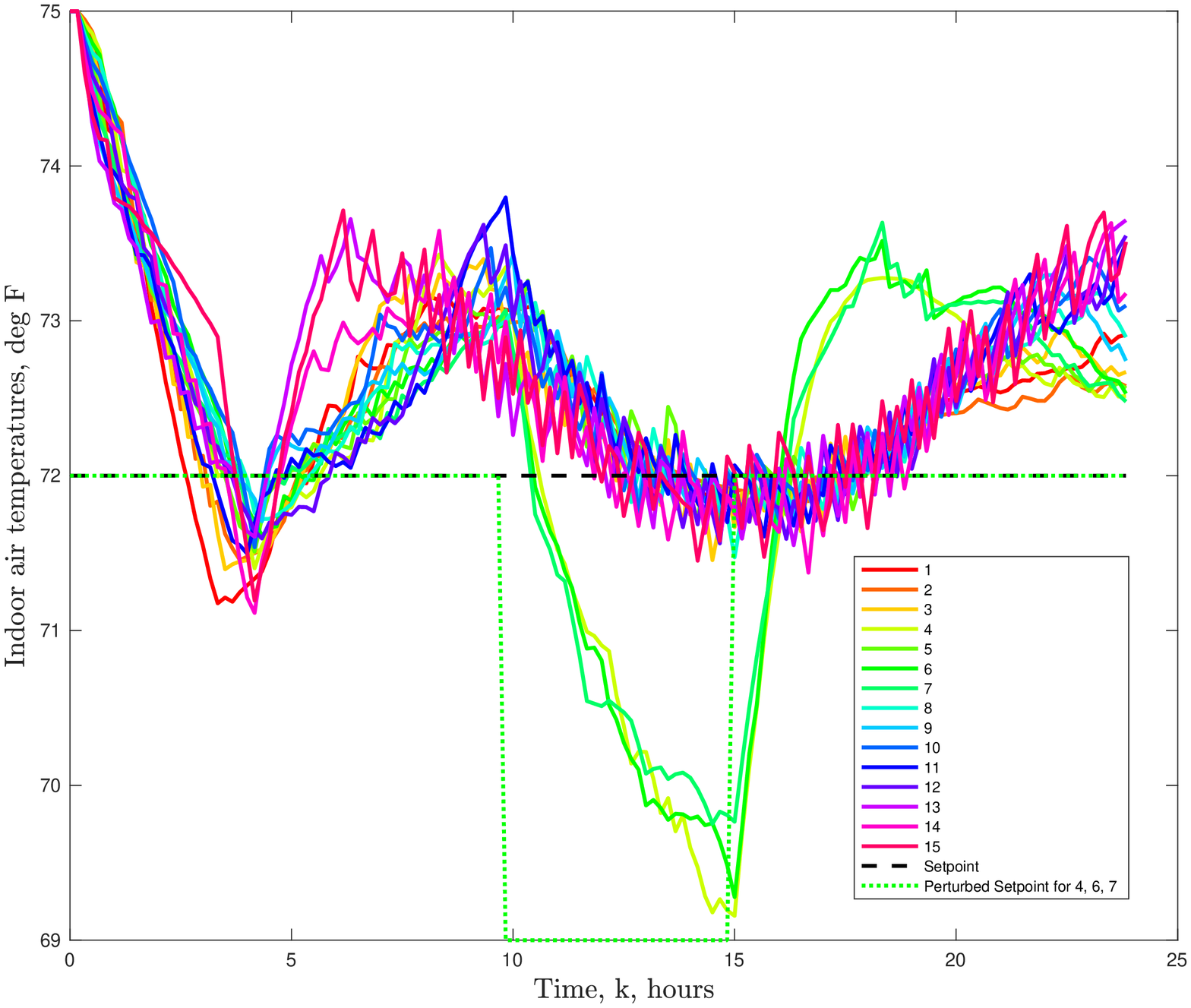}
    \caption{With swaps and local state feedback controller of 15 agents under disturbed setpoints}
    \label{fig:DynamicAllocation_N15Gaussian_LocalCtr_Swaps}
\end{figure}

The decentralized dynamic resource allocation solution proposed in this work follows the idea of \quotes{\textit{Global trendsetting, local negotiations}}. Here, the the global trend is for the agents' resources to be Gaussian distributed while summing up to the available power, and the local negotiations happen to maintain the balance between following such global trend and accounting for the local requirements simultaneously.

\section{Conclusions}
\label{sec::conclusions}

CVTs in one-dimensional spaces are desirable due to their inherent line structure and ease of computation of the entire tessellations, allowing for verification of the quality of the solution. Employing them in the resource allocation problem brings forth additional advantages as embedding desired global trends in the team through probability distributions. 

For a fixed amount of resource, the static resource allocation method offers an analytical, although central, solution by posing the constrained CVT minimization problem as a system of non-linear equations. The generalizability of the developed static allocation framework is worth remarking. Instead of the summation constraint, one can have any constraint from $\mathbb{R}^N \to \mathbb{R}$, and one can also have as many constraints as the number of parameters defining the desired distribution.

The developed decentralized dynamic allocation solution using CVTs provides a natural way to embed the aggregate team behavior or to set the desired global trend through the distribution of the tessellations. The developed method through the civility model allows for flexibility on the local end by absorbing and distributing disturbances throughout the team. We observed a hint of inherent privacy in the architecture through the highly dynamic communication network that nevertheless was a simple line graph at all times, demonstrating the scalability of the method.

Building on this work, we aim to generalize the develop decentralized resource allocation method to global trends that are described by more distributions, and not just Gaussian. We aim to verify the robustness of the architecture to a changing number of agents in the team, possibly due to communication failures. 

\bibliography{references}

\begin{thebibliography}{10}

\bibitem{GeneralCompetitiveAnalysis}
K.~Arrow and F.~H. Hahn, {\em General Competitive Analysis}.
\newblock North Holland, December 1983.

\bibitem{Hariharan}
H.~Lakshmanan and D.~P. de~Farias, ``Decentralized resource allocation in
  dynamic networks of agents,'' {\em SIAM Journal on Optimization}, vol.~19,
  no.~2, pp.~911--940, 2008.

\bibitem{ResourceAllocationBoyd}
L.~Xiao and S.~Boyd, ``Optimal scaling of a gradient method for distributed
  resource allocation,'' {\em Journal of Optimization Theory and Applications},
  vol.~129, no.~3, pp.~469--488, 2006.

\bibitem{ResourceAllocationADMMBanjac}
G.~Banjac, F.~Rey, P.~Goulart, and J.~Lygeros, ``Decentralized resource
  allocation via dual consensus {ADMM},'' in {\em 2019 American Control
  Conference (ACC)}, pp.~2789--2794, 2019.

\bibitem{ResourceAllocationUncertainty}
T.~T. Doan and C.~L. Beck, ``Distributed resource allocation over dynamic
  networks with uncertainty,'' {\em IEEE Transactions on Automatic Control},
  vol.~66, no.~9, pp.~4378--4384, 2021.

\bibitem{ResourceAllocationCai2012DecentralizedCO}
H.~Cai, ``Decentralized control of stochastic dynamic systems with applications
  to resource allocation and portfolio management,'' 2012.

\bibitem{JohnSnowCholera}
J.~Snow, {\em On the Mode of Communication of Cholera}.
\newblock 1855.

\bibitem{LiliJu2011}
L.~Ju, T.~Ringler, and M.~Gunzburger, {\em {V}oronoi Tessellations and Their
  Application to Climate and Global Modeling}, pp.~313--342.
\newblock Berlin, Heidelberg: Springer Berlin Heidelberg, 2011.

\bibitem{CortesMartinezMobileSensing}
J.~Cortes, S.~Martinez, T.~Karatas, and F.~Bullo, ``Coverage control for mobile
  sensing networks,'' {\em IEEE Transactions on Robotics and Automation},
  vol.~20, no.~2, pp.~243--255, 2004.

\bibitem{OptimalTransportCoverageControl}
D.~Inoue, Y.~Ito, and H.~Yoshida, ``Optimal transport-based coverage control
  for swarm robot systems: Generalization of the {V}oronoi tessellation-based
  method,'' {\em IEEE Control Systems Letters}, vol.~5, no.~4, pp.~1483--1488,
  2021.

\bibitem{AdaptiveConverageControl}
Y.~Bai, Y.~Wang, M.~Svinin, E.~Magid, and R.~Sun, ``Adaptive multi-agent
  coverage control with obstacle avoidance,'' {\em IEEE Control Systems
  Letters}, vol.~6, pp.~944--949, 2022.

\bibitem{SparsityStructureOptimalityCoverageControl}
A.~Davydov and Y.~Diaz-Mercado, ``Sparsity structure and optimality of
  multi-robot coverage control,'' {\em IEEE Control Systems Letters}, vol.~4,
  no.~1, pp.~13--18, 2020.

\bibitem{ElecMarketReformChen}
T.~Chen, H.~Pourbabak, and W.~Su, ``Electricity market reform,'' {\em The
  Energy Internet}, pp.~97--121, 2019.

\bibitem{BuiltEnvJournalWILLIAMS2020178}
S.~Williams and M.~Short, ``Electricity demand forecasting for decentralised
  energy management,'' {\em Energy and Built Environment}, vol.~1, no.~2,
  pp.~178 -- 186, 2020.

\bibitem{P2Preview}
C.~Park and T.~Yong, ``Comparative review and discussion on {P}2{P} electricity
  trading,'' {\em Energy Procedia}, vol.~128, pp.~3--9, 2017.

\bibitem{p2preview2}
C.~Zhang, J.~Wu, C.~Long, and M.~Cheng, ``Review of existing peer-to-peer
  energy trading projects,'' {\em Energy Procedia}, vol.~105, pp.~2563--2568,
  2017.

\bibitem{BEUCaggregators}
A.~Malizou, ``Electricity aggregators: Starting off on the right foot with
  consumers,'' {\em BEUC, The European Consumer Organization}, 2018.

\bibitem{aggregatorReview}
S.~Burger, J.~Chaves-\'{A}vila, C.~Batlle, and I.~P\'{e}rez-Arriaga, ``A review
  of the value of aggregators in electricity systems,'' {\em Renewable and
  Sustainable Energy Reviews}, vol.~77, pp.~395--405, 2017.

\bibitem{CVT_QiangDu}
Q.~Du, V.~Faber, and M.~Gunzburger, ``Centroidal voronoi tessellations:
  Applications and algorithms,'' {\em SIAM Review}, vol.~41, no.~4,
  pp.~637--676, 1999.

\bibitem{CVT_Fleischer}
P.~Fleischer, ``Sufficient conditions for achieving minimum distortion in a
  quantizer,'' {\em IEEE International Convention Record, Pt I}, pp.~104--111,
  1964.

\bibitem{UniquenessCVT_Urschel}
J.~C. Urschel, ``On the characterization and uniqueness of centroidal {V}oronoi
  tessellations,'' {\em SIAM Journal on Numerical Analysis}, vol.~55, no.~3,
  pp.~1525--1547, 2017.

\bibitem{CVT_Llyod_original}
S.~Lloyd, ``Least squares quantization in pcm,'' {\em IEEE Transactions on
  Information Theory}, vol.~28, no.~2, pp.~129--137, 1982.

\bibitem{ConvergenceLlyod_QiangDu}
Q.~Du, M.~Emelianenko, and L.~Ju, ``Convergence of the lloyd algorithm for
  computing centroidal voronoi tessellations,'' {\em SIAM Journal on Numerical
  Analysis}, vol.~44, no.~1, pp.~102--119, 2006.

\bibitem{CVT_1d_uniquenessKieffer}
J.~Kieffer, ``Uniqueness of locally optimal quantizer for log-concave density
  and convex error weighting function,'' {\em IEEE Transactions on Information
  Theory}, vol.~29, no.~1, pp.~42--47, 1983.

\bibitem{CVT_LloydsAlt_FastConvergenceLiu}
Y.~Liu, W.~Wang, B.~L\'{e}vy, F.~Sun, D.-M. Yan, L.~Lu, and C.~Yang, ``On
  centroidal {V}oronoi tessellation---energy smoothness and fast computation,''
  {\em ACM Trans. Graph.}, vol.~28, sep 2009.

\bibitem{CVT_LlyodsAlt_FastConvergenceWang}
X.~Wang, X.~Ying, Y.-J. Liu, S.-Q. Xin, W.~Wang, X.~Gu, W.~Mueller-Wittig, and
  Y.~He, ``Intrinsic computation of centroidal voronoi tessellation (cvt) on
  meshes,'' {\em Computer-Aided Design}, vol.~58, pp.~51--61, 2015.
\newblock Solid and Physical Modeling 2014.

\bibitem{CVT_LlyodsAlt_FastConvergenceHateley}
J.~Hatless, H.~Wei, and C.~L, ``Fast methods for computing centroidal voronoi
  tessellations,'' {\em Journal of Scientific Computing}, vol.~63,
  pp.~185--212, 2015.

\bibitem{JMacQueen}
J.~Macqueen, ``Some methods for classification and analysis of multivariate
  observations,'' {\em In 5-th Berkeley Symposium on Mathematical Statistics
  and Probability}, pp.~281--297, 1967.

\bibitem{GraphTheoryIntro}
M.~Loebl, {\em Introduction to Graph Theory}, pp.~13--49.
\newblock Wiesbaden: Vieweg+Teubner, 2010.

\bibitem{FredholmIntegralEquations}
A.-M. Wazwaz, {\em Fredholm Integral Equations}, pp.~119--173.
\newblock Berlin, Heidelberg: Springer Berlin Heidelberg, 2011.

\bibitem{XiaoMa_HVACmodel}
X.~Ma, J.~Dong, S.~M. Djouadi, J.~J. Nutaro, and T.~Kuruganti, ``Stochastic
  control of energy efficient buildings: A semidefinite programming approach,''
  in {\em 2015 IEEE International Conference on Smart Grid Communications
  (SmartGridComm)}, pp.~780--785, 2015.

\bibitem{KadirMFCpaper}
K.~Amasyali, Y.~Chen, B.~Telsang, M.~Olama, and S.~M. Djouadi, ``Hierarchical
  model-free transactional control of building loads to support grid
  services,'' {\em IEEE Access}, vol.~8, pp.~219367--219377, 2020.

\end{thebibliography}
\bibliographystyle{ieeetr}

\end{document}